\documentclass[amsmath,amssymb,preprint,prl,superscriptaddress]{revtex4}
\usepackage{amsfonts}
\usepackage{amsmath}
\usepackage{amsthm}
\usepackage{amscd}
\usepackage{amssymb}
\usepackage{subfigure}
\usepackage{amsxtra}
\usepackage{gensymb}
\usepackage{bm}           
\usepackage{bbm}
\usepackage{graphicx}
\usepackage{epstopdf}
\usepackage{color}
\usepackage{times}
\usepackage{mdframed}
\usepackage{textpos}
\usepackage[colorlinks=true,citecolor=blue,urlcolor=black]{hyperref}

\def\bi#1\ei {\begin{itemize}#1\end{itemize}}
\def\bn#1\en {\begin{enumerate}#1\end{enumerate}}
\def\bea#1\eea {\begin{align}#1\end{align}}
\def\bean#1\eean {\begin{align*}#1\end{align*}}
\def\ben#1\een {\begin{equation*}#1\end{equation*}}
\def\be#1\ee {\begin{equation}#1\end{equation}}
\def\bes#1\ees {\begin{equation}\begin{split}#1\end{split}\end{equation}}
\def\bear#1\eear {\begin{eqnarray}#1\end{eqnarray}}
\def\bear#1\eear {\begin{eqnarray*}#1\end{eqnarray*}}

\newcommand{\beq}{\begin{equation}}
\newcommand{\eeq}{\end{equation}}

\begin{document}

\title{\bf Color Erasure Detectors Enable Chromatic Interferometry}
\begin{textblock*}{5cm}(14.5cm,-1.5cm)
  \fbox{\footnotesize MIT-CTP-5111}
\end{textblock*}

\author{Luo-Yuan Qu}
\affiliation{Shanghai Branch, National Laboratory for Physical Sciences at Microscale and Department of Modern Physics University of Science and Technology of China, Shanghai, 201315, P.~R.~China}
\affiliation{CAS Center for Excellence and Synergetic Innovation Center in Quantum Information and Quantum Physics, Shanghai Branch,  University of Science and Technology of China, Shanghai, 201315, P.~R.~China}
\affiliation{Jinan Institute of Quantum Technology, Jinan, 250101, P.~R.~China}

\author{Jordan Cotler}
\affiliation{Stanford Institute for Theoretical Physics, Stanford University, Stanford, CA 94305 USA}

\author{Fei Ma}
\affiliation{Shanghai Branch, National Laboratory for Physical Sciences at Microscale and Department of Modern Physics University of Science and Technology of China, Shanghai, 201315, P.~R.~China}
\affiliation{CAS Center for Excellence and Synergetic Innovation Center in Quantum Information and Quantum Physics, Shanghai Branch,  University of Science and Technology of China, Shanghai, 201315, P.~R.~China}
\affiliation{Jinan Institute of Quantum Technology, Jinan, 250101, P.~R.~China}

\author{Jian-Yu Guan}
\affiliation{Shanghai Branch, National Laboratory for Physical Sciences at Microscale and Department of Modern Physics University of Science and Technology of China, Shanghai, 201315, P.~R.~China}
\affiliation{CAS Center for Excellence and Synergetic Innovation Center in Quantum Information and Quantum Physics, Shanghai Branch,  University of Science and Technology of China, Shanghai, 201315, P.~R.~China}

\author{Ming-Yang Zheng}
\author{Xiu-Ping Xie}
\affiliation{Jinan Institute of Quantum Technology, Jinan, 250101, P.~R.~China}

\author{Yu-Ao Chen}
\affiliation{Shanghai Branch, National Laboratory for Physical Sciences at Microscale and Department of Modern Physics University of Science and Technology of China, Shanghai, 201315, P.~R.~China}
\affiliation{CAS Center for Excellence and Synergetic Innovation Center in Quantum Information and Quantum Physics, Shanghai Branch,  University of Science and Technology of China, Shanghai, 201315, P.~R.~China}

\author{Qiang Zhang}
\affiliation{Shanghai Branch, National Laboratory for Physical Sciences at Microscale and Department of Modern Physics University of Science and Technology of China, Shanghai, 201315, P.~R.~China}
\affiliation{CAS Center for Excellence and Synergetic Innovation Center in Quantum Information and Quantum Physics, Shanghai Branch,  University of Science and Technology of China, Shanghai, 201315, P.~R.~China}
\affiliation{Jinan Institute of Quantum Technology, Jinan, 250101, P.~R.~China}

\author{Frank Wilczek}
\affiliation{Center for Theoretical Physics, MIT, Cambridge, MA 02139 USA}
\affiliation{T. D. Lee Institute, Shanghai Jiao Tong University, Shanghai, 200240, P.~R.~China}
\affiliation{Wilczek Quantum Center, School of Physics and Astronomy, Shanghai Jiao Tong University, Shanghai, 200240, P.~R.~China}
\affiliation{Department of Physics, Stockholm University, Stockholm SE-106 91 Sweden}
\affiliation{Department of Physics and Origins Project, Arizona State University, Tempe, AZ 25287 USA}

\author{Jian-Wei Pan}
\affiliation{Shanghai Branch, National Laboratory for Physical Sciences at Microscale and Department of Modern Physics University of Science and Technology of China, Shanghai, 201315, P.~R.~China}
\affiliation{CAS Center for Excellence and Synergetic Innovation Center in Quantum Information and Quantum Physics, Shanghai Branch,  University of Science and Technology of China, Shanghai, 201315, P.~R.~China}


\begin{abstract}
By engineering and manipulating quantum entanglement between incoming photons and experimental apparatus, we construct single-photon detectors which cannot distinguish between photons of very different wavelengths.  These color erasure detectors enable a new kind of intensity interferometry, with potential applications in microscopy and astronomy.  We demonstrate chromatic interferometry experimentally, observing robust interference using both coherent and incoherent photon sources.
\end{abstract}

\maketitle

Quantum interference ~\cite{hariharan2010basics}, which lies at the heart of quantum theory, requires complete indistinguishability between two particles. This is to say, as long as one can distinguish two particles even in principle, quantum interference will not happen. Meanwhile, quantum mechanics tells us if we can erase the two particles' past ~\cite{scully1982quantum}, interference will be restored. It has been shown that path or polarization information can be easily erased, while the frequency difference is generally hard to eliminate. For photons, conventional optical detectors are fundamentally photon counters, whose operation depends upon processes which are sensitive to the photons' energy. Thus, conventional detectors distinguish between different wavelengths, and therefore optical interference normally involves quasimonochromatic light ~\cite{mandel1995optical}. Yet relative phases between photons of different wavelengths potentially provide a rich source of information.
It is quite astonishing that, theoretically, the frequency information can be erased ~\cite{cotler2016entanglement}. Here, we leveraged frequency-space entanglement to develop color erasure detectors and achieve intensity interferometry ~\cite{brown1956test,baym1998physics} between light of very different wavelengths experimentally, thus revealing new features of optical radiation fields. This new type of interferometer might find immediate applications in astronomy, microscopy, and metrology ~\cite{monnier2003optical,shtengel2009interferometric,giovannetti2011advances}.

Since the final stage of optical detection generally involves quantized processes, i.e.~absorption or inelastic scattering, it is appropriate to use the language of photons.  Consider two sources $S_{1}$, $S_{2}$ which emit photons of different colors $\gamma_1$, $\gamma_2$ which are received at detectors $A$, $B$.  Simultaneous firing of $A$, $B$ can be achieved in two ways: $\gamma_1$ excites $A$ and $\gamma_2$ excites $B$, or $\gamma_2$ excites $A$ and $\gamma_1$ excites $B$.  If those two possibilities can be distinguished, then there is no interference between them. But if the detectors erase the color information, then interference will occur.  Let us emphasize that according to the principles of quantum theory, interference only occurs if the two final states are strictly indistinguishable.  Such strict color erasureness cannot be achieved simply by ignoring color information.  Rather, one must erase it. To do that we entangle the photons to the detectors using nonlinear processes ~\cite{cotler2016entanglement}.

In particular, we generate entanglement between an incoming $\gamma_1$ or $\gamma_2$ photon and a color erasure detector.  If the difference in energy between $\gamma_1$ and $\gamma_2$ is $\Delta E$, then a color erasure detector implements an entangling unitary of the form
\begin{equation}
\begin{aligned}
|\gamma_1\rangle |\text{detector}\rangle \longrightarrow &\frac{1}{\sqrt{2}} \, |\gamma_1\rangle |\text{detector, measured }\gamma_1\rangle + \frac{1}{\sqrt{2}}\, |\gamma_2\rangle |\text{detector} - \Delta E\,, \text{ measured }\gamma_2\rangle \\
|\gamma_2\rangle |\text{detector}\rangle \longrightarrow &-\frac{1}{\sqrt{2}} \, |\gamma_1\rangle |\text{detector} + \Delta E,\text{ measured }\gamma_1\rangle + \frac{1}{\sqrt{2}}\, |\gamma_2\rangle |\text{detector, measured }\gamma_2 \rangle
\end{aligned}
\end{equation}
where $|\text{detector}\rangle$ is the initial state of the detector, $|\text{detector, measured }\gamma_1\rangle$ is the state of the detector having measured a $\gamma_1$ photon, $|\text{detector} + \Delta E, \text{ measured }\gamma_1\rangle$ is the state of the detector having gained an energy $\Delta E$ and also having measured a $\gamma_1$ photon, and the other states are defined similarly.  If we only consider occurrences where $\gamma_1$ is measured (i.e., project onto final states with a $|\gamma_1\rangle$), then we are left with either
\begin{equation}
\begin{aligned}
\label{eq:finalstates1}
|\gamma_1\rangle |\text{detector, measured }\gamma_1\rangle\quad \text{    or    }  \quad |\gamma_1\rangle |\text{detector} + \Delta E,\text{ measured }\gamma_1\rangle
\end{aligned}
\end{equation}
the first state having come from an initial $\gamma_1$ photon and the second state having come from an initial $\gamma_2$ photon.  The key point is that the overlap of the final detector states is approximately $1$, namely
\begin{equation}
\langle  \text{detector, measured }\gamma_1 | \text{detector} + \Delta E,\text{ measured }\gamma_1\rangle \approx 1\,,
\end{equation}
and so our two final states in Eqn.~\eqref{eq:finalstates1} are essentially indistinguishable, regardless of whether the initial incoming photon was $\gamma_1$ or $\gamma_2$.  In other words, by generating a specific kind of entangled state between the incoming photon and the detector, we can cause decoherence (via our projective measurement) to quantum mechanically erase the color information of the initial photon.  More details about these entangled states can be found in the Supplementary Materials.

Our color erasure detectors are technically and conceptually distinct from previous experiments in frequency-space interferometry.  Conventional interferometry experiments, such as Mach-Zehnder and Hong-Ou-Mandel interferometry, are performed with standard beamsplitters, but can equally well be performed with light beams of distinct polarization and polarizing beamsplitters.  In this spirit, recently more sophisticated experiments ~\cite{kobayashi2016frequency,kobayashi2017mach} performed Mach-Zehnder and Hong-Ou-Mandel interferometry with light beams of distinct frequency and frequency-space beamsplitters.  By contrast, color erasure detectors retroactively recover interference from conventional interferometry experiments performed with standard beamsplitters but distinct frequencies of light.  This is akin to quantum eraser experiments ~\cite{scully1991quantum,kwiat1992observation}, but now involving erasure of color information.  An important advantage of our approach is that only detection apparatus requires augmentation.  This is convenient in general, and essential for imaging tasks involving self-luminous sources.

\begin{figure*}[tbh]
\centering
\resizebox{14cm}{!}{\includegraphics{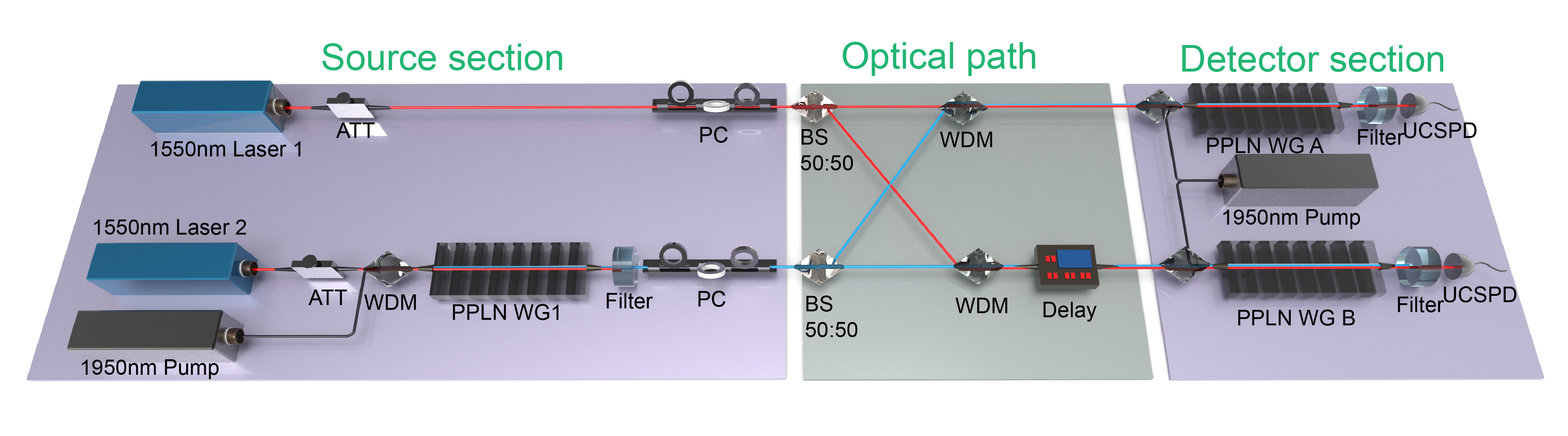}}
\caption{\textbf{Diagram of the chromatic intensity interferometer.} VOA: variable optical attenuator, PPLN: periodically-poled lithium niobate, BPF: an 863 nm band pass filter, PC: polarization controller, BS: beamsplitters, WDMs: two wavelength division multiplexers, UCSPDs: upconversion single photon detectors.}
\label{fig1}
\end{figure*}

We realize chromatic intensity interferometry with our color erasure detectors. As shown in Fig.~1, we first choose an attenuated $1550$ nm laser as the source of $\gamma_1$. With the help of an $1950$ nm pump laser, we up-convert another independent $1550$ nm laser light into $863$ nm light via sum-frequency generation (SFG) in a home-made straight periodically-poled lithium niobate (PPLN) waveguide ~\cite{zheng2016integrated} (PPLN WG1). An $863$ nm band pass filter is exploited to block the 1950 nm pump and the $863$ nm light is taken as the source of $\gamma_2$. We then use beamsplitters and wavelength division multiplexing (WDM) to divide and couple photons from both sources to the color erasure detectors, which are composed of two integrated PPLN waveguides (PPLN WG A,B) ~\cite{ma2018upconversion}, a $1950$ nm pump source, band pass filters, and two telecom band single photon detectors ~\cite{zheng2016integrated}.

In order to observe color erasure interference, we need to change the relative phase between the $\gamma_1$ and $\gamma_2$ photons in one arm of the detector ~\cite{cotler2016entanglement}.  Since the phase of a $\gamma_2$ photon changes faster than that of a $\gamma_1$ photon with the same delay time, we can control the relative phase by adjusting the optical fiber delay (MDL-002) before detector $B$.  We can choose the final output of the color erasure detectors to be either $\gamma_1$ or $\gamma_2$, contingent on our choice of band pass filters.  We record the arrival time of each photon by a time-digital converter (TDC) and a computer.

Generally, intensity interferometry is observed in terms of $g^{(2)}(\tau)$, the second-order quantum mechanical correlation function. As we can see in the red curve in Fig.~2(a), the correlation $g^{(2)}(\tau = 3\text{ ns})$ oscillates as we change the optical delay and detect $\gamma_1$ photons by filtering out the $\gamma_2$ photons.  Photons from lasers obey Poissonian number statistics so that the $\tau$-average of $g^{(2)}(\tau)$ is $1$.

\begin{figure}[h!]
\centering
\resizebox{9cm}{!}{\includegraphics{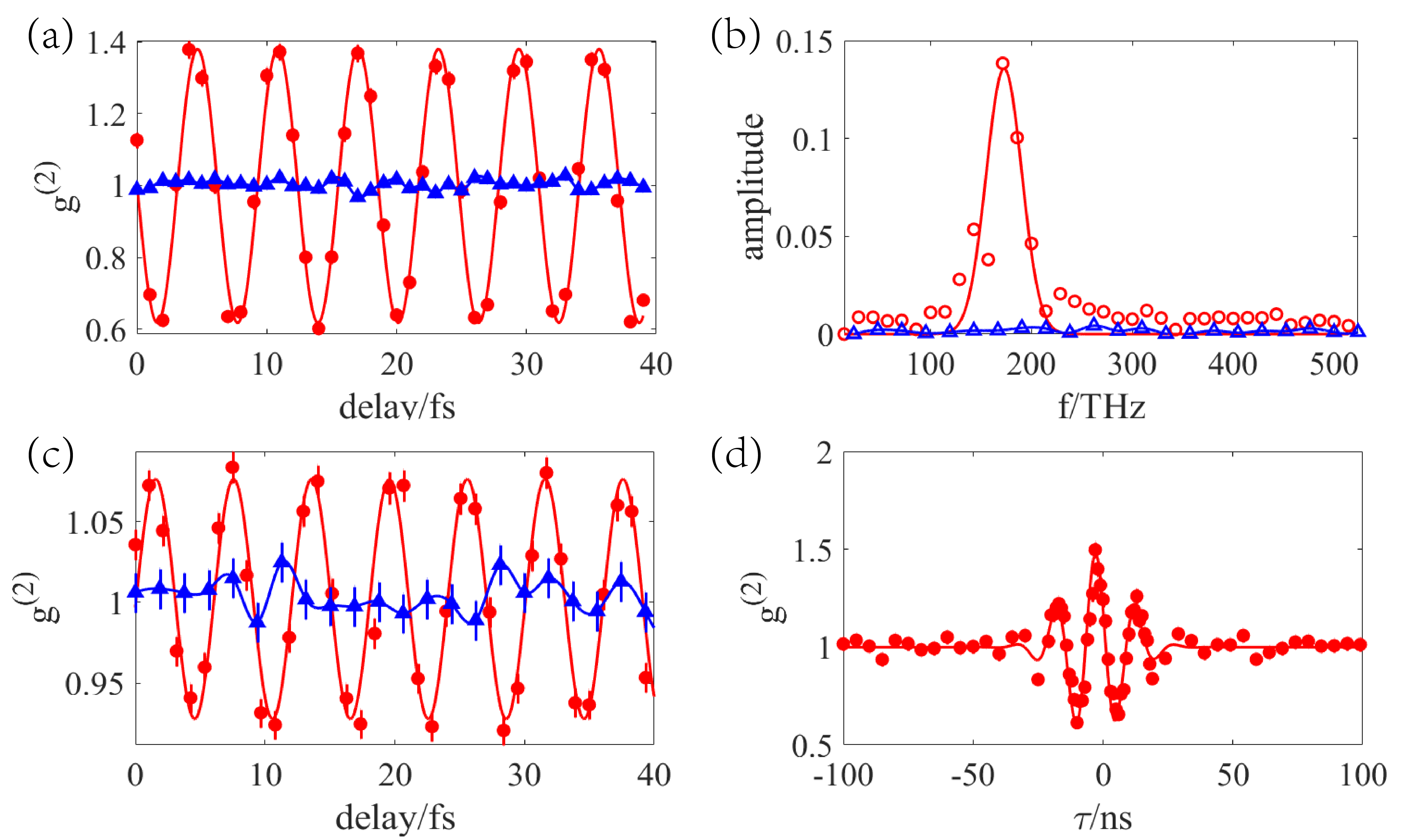}}
\caption{\textbf{Chromatic intensity interferometry of lasers.} (a) $g^{(2)}(\tau = 3\text{ ns})$ as a function of the optical delay time, where the color erasure detectors each output 1550 nm light. The red rounded markers display interference of different wavelengths of light due to the color erasure detectors, whereas the blue triangle markers do not display interference since standard detectors are used.  The same color scheme is used in (b) and (c). (b) The Fourier transform of $g^{(2)}(\tau = 3\text{ ns})$ as a function of the optical delay time.  (c) $g^{(2)}(\tau = 3\text{ ns})$ as a function of the optical delay time, where the color erasure detectors each output 863 nm light. (d) $g^{(2)}(\tau)$ as a function of the delay $\tau$ between the two detectors.}
\label{fig2}
\end{figure}

The visibility of the interference is around $0.4$, slightly less than the theoretically expected visibility $0.5$ mainly due to the up-conversion single photon detector's dark counts and baseline error from imperfect devices. For comparison, we also measure $g^{(2)}(\tau)$ without the pump light which enables the detectors to distinguish between the incoming wavelengths, so they are no longer color erasure.  As expected, the interference pattern disappears, as shown by the blue curve of Fig.~2(a).

Fig.~2(b) shows the Fourier transforms of the two curves in Fig.~2(a). The location of the peak of the red curve represents the frequency of the interference pattern, i.e. the rate of phase change as we scan the optical delay. In our case, the rate of phase change is theoretically the frequency of pump. The measured peak position is around $144$ THz, which well-coincides with $1950$ nm. The blue curve in Fig.~2(b) is just noise and so has no large peaks, demonstrating that interference does not occur in the absence of color erasure detectors.

Instead of having each color erasure detector output 1550 nm light, we can instead arrange that the detectors each output 863 nm light.  Data for this alternative arrangement are shown in Fig.~2(c).  In the figure, we filter in only $\gamma_2$ photons at the output of the waveguides, and collect coincidence counts with and without the pumps enabling color erasure detection.  Relative to filtering in $\gamma_1$ photons, the visibility of interference when filtering in $\gamma_2$ photons is degraded since the photons tend to be multi-mode when propagating through the PPLN waveguides comprising our color erasure detectors.  Only photons in the lowest transverse mode participate in interference. The photons in other modes induce noise and thus reduce the visibility.

We also perform Hong-Ou-Mandel interference  ~\cite{hong1987measurement} utilizing standard beamsplitters and two different wavelengths of light.  The interference can only be recovered with color erasure detectors.  Instead of changing the relative time delay of the light beams, we instead observe coincidence counts between different time slots in the TDC.  In Fig.~2(d), we observe an oscillation of $g^{(2)}(\tau)$ as a function of $\tau$, which decays as the delay between two detectors surpasses the coherence time of the light sources. We can produce bunching or antibunching depending on the setup of the interferometer, and the settings of the color erasure detectors.

In a tabletop demonstration experiment, it is convenient to use lasers as light sources. Considering future applications, we would like to observe chromatic interferometry for incoherent or semi-incoherent sources such as thermal light from a star or photon emission from fluorescent proteins.  Therefore, it is important to demonstrate that our chromatic intensity interferometer can function with thermal light.  Accordingly, we experimentally performed chromatic intensity interferometry with thermal light sources.  To construct a thermal source, we prepare a C band amplified spontaneous emission (ASE) light source with 30 nm spectral bandwidth. We first filter the ASE light with a 100 GHz bandwidth dense wavelength division multiplexer (DWDM) and then amplify it with an Erbium doped fiber amplifier (EDFA). The emission of EDFA is further filtered by a 50 MHz bandwidth etalon to select out a thermally populated mode which is then divided into two beams. One is used for $\gamma_1$ and the other one is converted to $863$ nm in a PPLN waveguide to become $\gamma_2$, similar to the coherent laser setting from before.  In this thermal source setup, the $\gamma_1$ and $\gamma_2$ photons are generated from the same source and thus their phases are correlated.  To destroy these correlations, the $\gamma_1$ beam is sent through a 20 km spool of fiber, and fluctuations of the fiber ruin the phase coherence between $\gamma_1$ and $\gamma_2$. Then we send both beams to the color erasure detectors and observe interference.

\begin{figure}[h!]
\centering
\resizebox{9cm}{!}{\includegraphics{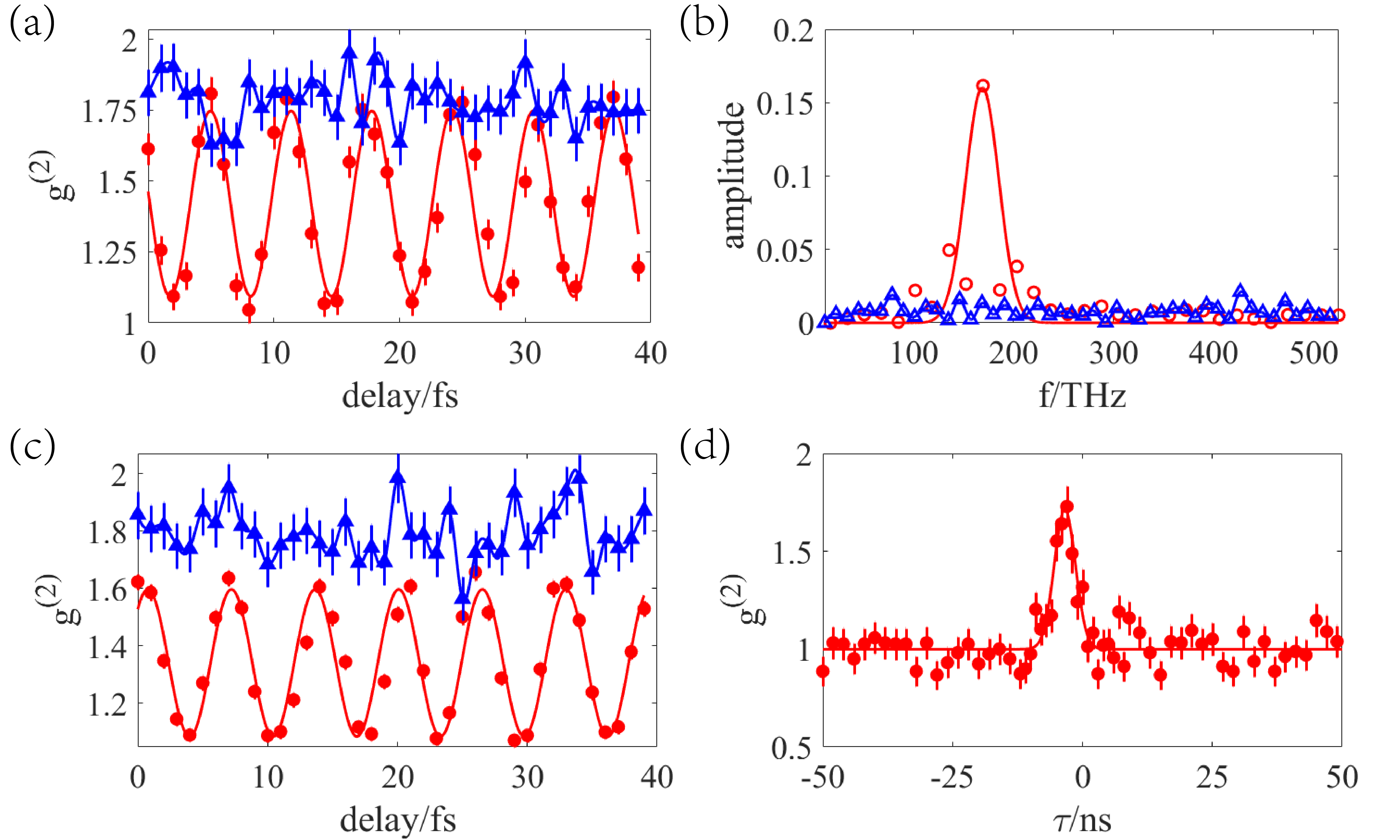}}
\caption{\textbf{Chromatic intensity interferometry of thermal sources.} (a) $g^{(2)}(\tau)$ for $\tau \approx 0$ as a function of the optical delay time, where the color erasure detectors each output 1550 nm light. The red rounded markers show interference due to the aid of the color erasure detectors, whereas the blue triangle markers show the null outcome in the absence of color erasure detectors.  This color scheme is also used in (b) and (c). (b) The Fourier transform of $g^{(2)}(\tau)$ for $\tau \approx 0$ as a function of the optical delay time.  (c) $g^{(2)}(\tau)$ for $\tau \approx 0$ as a function of the optical delay time, where the color erasure detectors each output 863 nm light. (d) $g^{(2)}(\tau)$ as a function of the delay $\tau$ between the two detectors.}
\label{fig3}
\end{figure}

As shown in Fig.~3(a) and Fig.~3(c), we observed interference of the thermal light in when the color erasure detectors output only $|\gamma_1 \gamma_1 \rangle$ or only $|\gamma_2 \gamma_2\rangle$, respectively. We also compute the Fourier transform of the interference pattern for the $|\gamma_1 \gamma_1\rangle$ case. In the absence of color erasure detectors (i.e., by not pumping the waveguides), we check that interference does not occur.  We have also performed chromatic Hong-Ou-Mandel interferometry with these thermal sources, and $g^{(2)}(\tau)$ is shown in Fig.~3(d).

One apparent difference between our experimental data for thermal sources versus coherent lasers is the mean value of the interference patterns. In Fig.~3(a) and Fig.~3(c), the mean value is larger than $1$, which coincidences with the super-Poissonian number statistics of thermal light. The visibility for the thermal sources is worse than for the coherent lasers since the coherence time of the thermal sources is much shorter.  Thus every mismatch in the optical path will lead to the loss of coherence and visibility.

Since we expect color erasure detectors to have applications in free space imaging, we also performed chromatic interferometry in free space. As shown in Fig.~4(a), we detect the photons from two disk-like sources emitting different wavelengths of light.
\begin{figure}[h!]
\centering
\resizebox{9cm}{!}{\includegraphics{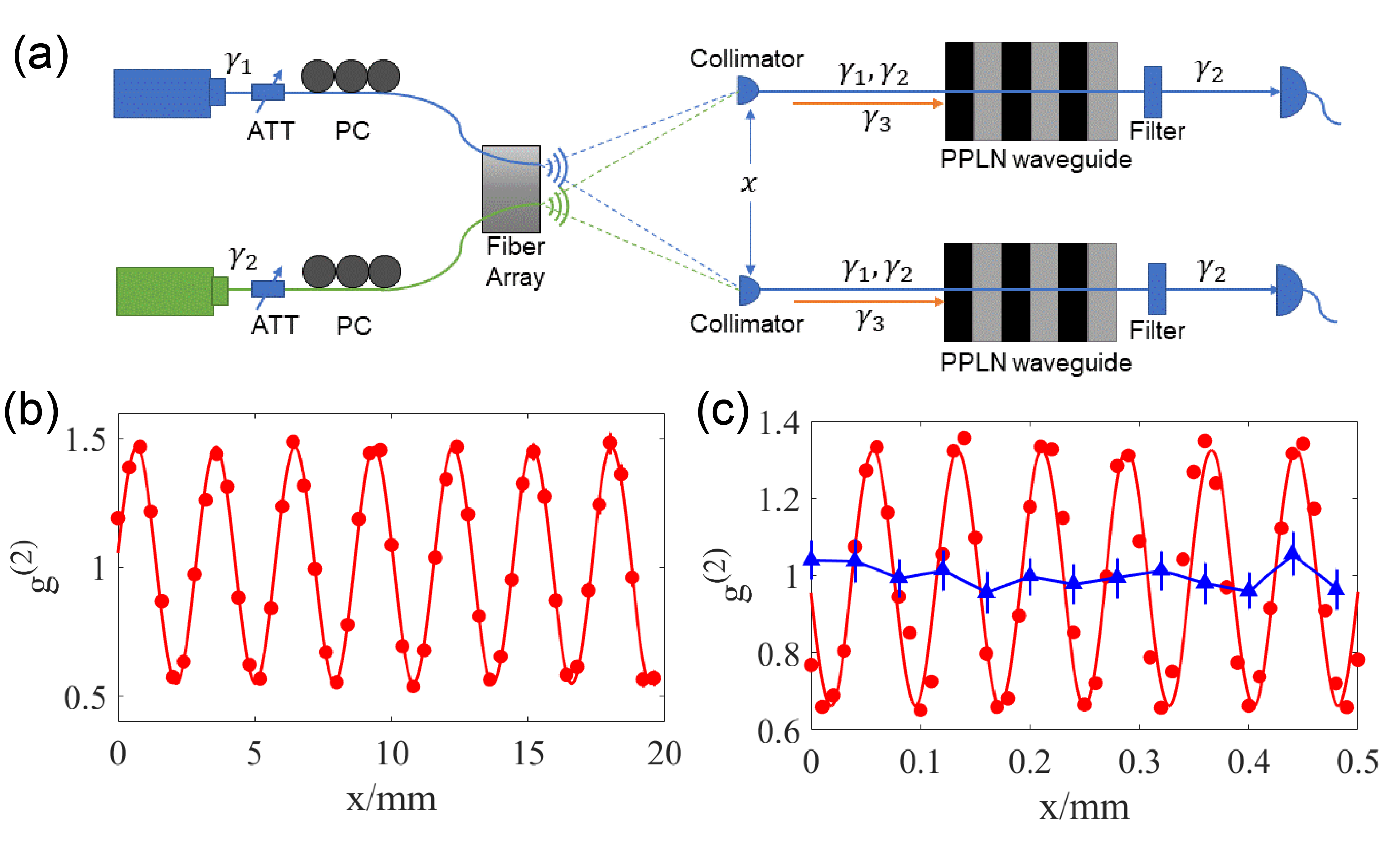}}
\caption{\textbf{Chromatic intensity interferometry in free space.} (a) A diagram of the experimental setup for the intensity interferometer in free space.   Lasers from a fiber array are utilized as sources, and the collimators are utilized to couple light from free space into color erasure detectors. One of the collimators is mounted on a linear translation stage to control the distance $x$ between the two collimators. (b) The measured interference pattern when both sources emit light of the same wavelength, reproducing standard Hanbury Brown and Twiss interference. (c) The measured interference pattern when the two sources emit 1550 nm and 863 nm light, respectively. The red rounded markers show interference due to the aid of the color erasure detectors, whereas the blue triangle markers show the absence of interference without the color erasure detectors. }
\label{fig4}
\end{figure}  

The disk-like sources are situated $125$ $\mu$m apart in a fiber array, and color erasure detectors are placed $40$ cm away. When we move the position of one of the detectors using a linear translation stage, we observe an interference pattern, as shown in the red curve in Fig.~4(c). The blue curve in Fig.~4(c) illustrates that interference is not observed in the absence of color erasure detection.  We also show in Fig.~4(b) the standard Hanbury Brown and Twiss interference pattern when the two sources emit at the same wavelength, utilizing standard detection apparatus. Our free space results for chromatic interferometry demonstrate the potential application of color erasure detection in imaging.

In conclusion, we have used our color erasure detectors to perform intensity interferometry between photons of very different wavelengths, and to recover their relative phase information, which is inaccessible to conventional detectors.  Since our technique does not require lenses, it could be used with very large apertures, and in regions of the spectrum where lenses are not readily available. This might inspire new opportunities for imaging and thus calls for further theoretical and experimental research.  As an example, color erasure detectors can enhance the ability of fluorescent microscopes  ~\cite{grussmayer2014photon,schwartz2013superresolution,leung2011review} to resolve nearby proteins which emit at distinct frequencies.  We can also leverage a generalization of the van Cittert-Zernike formula for sources of different wavelengths measured with color erasure detectors ~\cite{cotler2016entanglement}.

If instead we had a nearly perfect single photon detector, which has no noise, no jitter, no dead time and is very fast, we can effectively erase the frequency of incoming photons and use it in the multi-color HBT interferometer. However, there does not exist a photon detector or traditional photodiode faster than 144 THz, as would be required in our experiment. What's more, a fast detector acts like a very narrow timing filter, which filters the two input light pulses into a very narrow time window. This would filter out most of the photons in the pulses. In our experiment, the linewidth for the input laser is around 3 MHz and the detector bandwidth is around 144 THz. Only around 0.002\% (3 MHz/144THz) of the light will be detected. In this sense, it is indeed inefficient. Meanwhile, our system can convert photons with an efficiency of around 50\% which is orders of magnitude higher than a fast detection method. This is actually not due to a technological advance but a difference in concept. Instead of filtering light, we coherently convert different wavelengths of light to become indistinguishable.

Our work exploits and emphasizes the realization that detectors are themselves quantum mechanical objects, which ``measure'' other systems by becoming entangled with them ~\cite{zurek2003decoherence,zurek2009quantum,cotler2016entanglement}.  Indeed, the core mechanism enabling multi-wavelength intensity interferometry is a trade-off between coherence of multi-photon phase information and coherence of color information, implemented by crafting and manipulating the entanglement between source photons and the detection apparatus.  (For mathematical details, see the Supplementary Materials.)  We anticipate that further analysis of the quantum mechanics of detectors will reveal other trade-off opportunities. 
\\ \\
\textbf{Acknowledgements} \\ \\
We would like to thank Andreas Kaldun and Philip Bucksbaum for valuable conversations and thank Lian-Tuan Xiao and Jian-Yong Hu for lending equipments. This work was supported by the National Key R\&D Program of China (2018YFB0504300), the National Natural Science Foundation of China, and the Chinese Academy of Science. JC is supported by the Fannie and John Hertz Foundation and the Stanford Graduate Fellowship program. FW’s work is supported by the the Swedish Research Council under Contract No. 335- 2014-7424, the U.S. Department of Energy under grant Contract No. DE-SC0012567, and by the European Research Council under grant 742104.

L.-Y. Q., J. C., and M. F. contributed equally to this work.

\bibliography{Bibliography.bib}

\begin{thebibliography}{28}
\expandafter\ifx\csname natexlab\endcsname\relax\def\natexlab#1{#1}\fi
\expandafter\ifx\csname bibnamefont\endcsname\relax
  \def\bibnamefont#1{#1}\fi
\expandafter\ifx\csname bibfnamefont\endcsname\relax
  \def\bibfnamefont#1{#1}\fi
\expandafter\ifx\csname citenamefont\endcsname\relax
  \def\citenamefont#1{#1}\fi
\expandafter\ifx\csname url\endcsname\relax
  \def\url#1{\texttt{#1}}\fi
\expandafter\ifx\csname urlprefix\endcsname\relax\def\urlprefix{URL }\fi
\providecommand{\bibinfo}[2]{#2}
\providecommand{\eprint}[2][]{\url{#2}}

\bibitem[{\citenamefont{Hariharan}(2010)}]{hariharan2010basics}
\bibinfo{author}{\bibfnamefont{P.}~\bibnamefont{Hariharan}},
  \emph{\bibinfo{title}{Basics of interferometry}}
  (\bibinfo{publisher}{Elsevier}, \bibinfo{year}{2010}).

\bibitem[{\citenamefont{Scully and Dr{\"u}hl}(1982)}]{scully1982quantum}
\bibinfo{author}{\bibfnamefont{M.~O.} \bibnamefont{Scully}} \bibnamefont{and}
  \bibinfo{author}{\bibfnamefont{K.}~\bibnamefont{Dr{\"u}hl}},
  \bibinfo{journal}{Physical Review A} \textbf{\bibinfo{volume}{25}},
  \bibinfo{pages}{2208} (\bibinfo{year}{1982}).

\bibitem[{\citenamefont{Mandel and Wolf}(1995)}]{mandel1995optical}
\bibinfo{author}{\bibfnamefont{L.}~\bibnamefont{Mandel}} \bibnamefont{and}
  \bibinfo{author}{\bibfnamefont{E.}~\bibnamefont{Wolf}},
  \emph{\bibinfo{title}{Optical coherence and quantum optics}}
  (\bibinfo{publisher}{Cambridge University Press}, \bibinfo{year}{1995}).

\bibitem[{\citenamefont{Cotler et~al.}(2016)\citenamefont{Cotler, Wilczek, and
  Borish}}]{cotler2016entanglement}
\bibinfo{author}{\bibfnamefont{J.}~\bibnamefont{Cotler}},
  \bibinfo{author}{\bibfnamefont{F.}~\bibnamefont{Wilczek}}, \bibnamefont{and}
  \bibinfo{author}{\bibfnamefont{V.}~\bibnamefont{Borish}},
  \bibinfo{journal}{arXiv preprint arXiv:1607.05719}  (\bibinfo{year}{2016}).

\bibitem[{\citenamefont{Brown and Twiss}(1956)}]{brown1956test}
\bibinfo{author}{\bibfnamefont{R.~H.} \bibnamefont{Brown}} \bibnamefont{and}
  \bibinfo{author}{\bibfnamefont{R.}~\bibnamefont{Twiss}},
  \bibinfo{journal}{Nature} \textbf{\bibinfo{volume}{178}},
  \bibinfo{pages}{1046} (\bibinfo{year}{1956}).

\bibitem[{\citenamefont{Baym}(1998)}]{baym1998physics}
\bibinfo{author}{\bibfnamefont{G.}~\bibnamefont{Baym}}, \bibinfo{journal}{Acta
  Physica Polonica. Series B} \textbf{\bibinfo{volume}{29}},
  \bibinfo{pages}{1839} (\bibinfo{year}{1998}).

\bibitem[{\citenamefont{Monnier}(2003)}]{monnier2003optical}
\bibinfo{author}{\bibfnamefont{J.~D.} \bibnamefont{Monnier}},
  \bibinfo{journal}{Reports on Progress in Physics}
  \textbf{\bibinfo{volume}{66}}, \bibinfo{pages}{789} (\bibinfo{year}{2003}).

\bibitem[{\citenamefont{Shtengel et~al.}(2009)\citenamefont{Shtengel,
  Galbraith, Galbraith, Lippincott-Schwartz, Gillette, Manley, Sougrat,
  Waterman, Kanchanawong, Davidson et~al.}}]{shtengel2009interferometric}
\bibinfo{author}{\bibfnamefont{G.}~\bibnamefont{Shtengel}},
  \bibinfo{author}{\bibfnamefont{J.~A.} \bibnamefont{Galbraith}},
  \bibinfo{author}{\bibfnamefont{C.~G.} \bibnamefont{Galbraith}},
  \bibinfo{author}{\bibfnamefont{J.}~\bibnamefont{Lippincott-Schwartz}},
  \bibinfo{author}{\bibfnamefont{J.~M.} \bibnamefont{Gillette}},
  \bibinfo{author}{\bibfnamefont{S.}~\bibnamefont{Manley}},
  \bibinfo{author}{\bibfnamefont{R.}~\bibnamefont{Sougrat}},
  \bibinfo{author}{\bibfnamefont{C.~M.} \bibnamefont{Waterman}},
  \bibinfo{author}{\bibfnamefont{P.}~\bibnamefont{Kanchanawong}},
  \bibinfo{author}{\bibfnamefont{M.~W.} \bibnamefont{Davidson}},
  \bibnamefont{et~al.}, \bibinfo{journal}{Proceedings of the National Academy
  of Sciences} \textbf{\bibinfo{volume}{106}}, \bibinfo{pages}{3125}
  (\bibinfo{year}{2009}).

\bibitem[{\citenamefont{Giovannetti et~al.}(2011)\citenamefont{Giovannetti,
  Lloyd, and Maccone}}]{giovannetti2011advances}
\bibinfo{author}{\bibfnamefont{V.}~\bibnamefont{Giovannetti}},
  \bibinfo{author}{\bibfnamefont{S.}~\bibnamefont{Lloyd}}, \bibnamefont{and}
  \bibinfo{author}{\bibfnamefont{L.}~\bibnamefont{Maccone}},
  \bibinfo{journal}{Nature Photonics} \textbf{\bibinfo{volume}{5}},
  \bibinfo{pages}{222} (\bibinfo{year}{2011}).

\bibitem[{\citenamefont{Kobayashi et~al.}(2016)\citenamefont{Kobayashi, Ikuta,
  Yasui, Miki, Yamashita, Terai, Yamamoto, Koashi, and
  Imoto}}]{kobayashi2016frequency}
\bibinfo{author}{\bibfnamefont{T.}~\bibnamefont{Kobayashi}},
  \bibinfo{author}{\bibfnamefont{R.}~\bibnamefont{Ikuta}},
  \bibinfo{author}{\bibfnamefont{S.}~\bibnamefont{Yasui}},
  \bibinfo{author}{\bibfnamefont{S.}~\bibnamefont{Miki}},
  \bibinfo{author}{\bibfnamefont{T.}~\bibnamefont{Yamashita}},
  \bibinfo{author}{\bibfnamefont{H.}~\bibnamefont{Terai}},
  \bibinfo{author}{\bibfnamefont{T.}~\bibnamefont{Yamamoto}},
  \bibinfo{author}{\bibfnamefont{M.}~\bibnamefont{Koashi}}, \bibnamefont{and}
  \bibinfo{author}{\bibfnamefont{N.}~\bibnamefont{Imoto}},
  \bibinfo{journal}{Nature Photonics} \textbf{\bibinfo{volume}{10}},
  \bibinfo{pages}{441} (\bibinfo{year}{2016}).

\bibitem[{\citenamefont{Kobayashi et~al.}(2017)\citenamefont{Kobayashi,
  Yamazaki, Matsuki, Ikuta, Miki, Yamashita, Terai, Yamamoto, Koashi, and
  Imoto}}]{kobayashi2017mach}
\bibinfo{author}{\bibfnamefont{T.}~\bibnamefont{Kobayashi}},
  \bibinfo{author}{\bibfnamefont{D.}~\bibnamefont{Yamazaki}},
  \bibinfo{author}{\bibfnamefont{K.}~\bibnamefont{Matsuki}},
  \bibinfo{author}{\bibfnamefont{R.}~\bibnamefont{Ikuta}},
  \bibinfo{author}{\bibfnamefont{S.}~\bibnamefont{Miki}},
  \bibinfo{author}{\bibfnamefont{T.}~\bibnamefont{Yamashita}},
  \bibinfo{author}{\bibfnamefont{H.}~\bibnamefont{Terai}},
  \bibinfo{author}{\bibfnamefont{T.}~\bibnamefont{Yamamoto}},
  \bibinfo{author}{\bibfnamefont{M.}~\bibnamefont{Koashi}}, \bibnamefont{and}
  \bibinfo{author}{\bibfnamefont{N.}~\bibnamefont{Imoto}},
  \bibinfo{journal}{Optics Express} \textbf{\bibinfo{volume}{25}},
  \bibinfo{pages}{12052} (\bibinfo{year}{2017}).

\bibitem[{\citenamefont{Scully et~al.}(1991)\citenamefont{Scully, Englert, and
  Walther}}]{scully1991quantum}
\bibinfo{author}{\bibfnamefont{M.~O.} \bibnamefont{Scully}},
  \bibinfo{author}{\bibfnamefont{B.-G.} \bibnamefont{Englert}},
  \bibnamefont{and} \bibinfo{author}{\bibfnamefont{H.}~\bibnamefont{Walther}},
  \bibinfo{journal}{Nature} \textbf{\bibinfo{volume}{351}},
  \bibinfo{pages}{111} (\bibinfo{year}{1991}).

\bibitem[{\citenamefont{Kwiat et~al.}(1992)\citenamefont{Kwiat, Steinberg, and
  Chiao}}]{kwiat1992observation}
\bibinfo{author}{\bibfnamefont{P.~G.} \bibnamefont{Kwiat}},
  \bibinfo{author}{\bibfnamefont{A.~M.} \bibnamefont{Steinberg}},
  \bibnamefont{and} \bibinfo{author}{\bibfnamefont{R.~Y.} \bibnamefont{Chiao}},
  \bibinfo{journal}{Physical Review A} \textbf{\bibinfo{volume}{45}},
  \bibinfo{pages}{7729} (\bibinfo{year}{1992}).

\bibitem[{\citenamefont{Zheng et~al.}(2016)\citenamefont{Zheng, Shentu, Ma,
  Zhou, Zhang, Dai, Xie, Zhang, and Pan}}]{zheng2016integrated}
\bibinfo{author}{\bibfnamefont{M.-Y.} \bibnamefont{Zheng}},
  \bibinfo{author}{\bibfnamefont{G.-L.} \bibnamefont{Shentu}},
  \bibinfo{author}{\bibfnamefont{F.}~\bibnamefont{Ma}},
  \bibinfo{author}{\bibfnamefont{F.}~\bibnamefont{Zhou}},
  \bibinfo{author}{\bibfnamefont{H.-T.} \bibnamefont{Zhang}},
  \bibinfo{author}{\bibfnamefont{Y.-Q.} \bibnamefont{Dai}},
  \bibinfo{author}{\bibfnamefont{X.}~\bibnamefont{Xie}},
  \bibinfo{author}{\bibfnamefont{Q.}~\bibnamefont{Zhang}}, \bibnamefont{and}
  \bibinfo{author}{\bibfnamefont{J.-W.} \bibnamefont{Pan}},
  \bibinfo{journal}{Review of Scientific Instruments}
  \textbf{\bibinfo{volume}{87}}, \bibinfo{pages}{093115}
  (\bibinfo{year}{2016}).

\bibitem[{\citenamefont{Ma et~al.}(2018)\citenamefont{Ma, Liang, Chen, Gao,
  Zheng, Xie, Liu, Zhang, and Pan}}]{ma2018upconversion}
\bibinfo{author}{\bibfnamefont{F.}~\bibnamefont{Ma}},
  \bibinfo{author}{\bibfnamefont{L.-Y.} \bibnamefont{Liang}},
  \bibinfo{author}{\bibfnamefont{J.-P.} \bibnamefont{Chen}},
  \bibinfo{author}{\bibfnamefont{Y.}~\bibnamefont{Gao}},
  \bibinfo{author}{\bibfnamefont{M.-Y.} \bibnamefont{Zheng}},
  \bibinfo{author}{\bibfnamefont{X.-P.} \bibnamefont{Xie}},
  \bibinfo{author}{\bibfnamefont{H.}~\bibnamefont{Liu}},
  \bibinfo{author}{\bibfnamefont{Q.}~\bibnamefont{Zhang}}, \bibnamefont{and}
  \bibinfo{author}{\bibfnamefont{J.-W.} \bibnamefont{Pan}},
  \bibinfo{journal}{JOSA B} \textbf{\bibinfo{volume}{35}},
  \bibinfo{pages}{2096} (\bibinfo{year}{2018}).

\bibitem[{\citenamefont{Hong et~al.}(1987)\citenamefont{Hong, Ou, and
  Mandel}}]{hong1987measurement}
\bibinfo{author}{\bibfnamefont{C.-K.} \bibnamefont{Hong}},
  \bibinfo{author}{\bibfnamefont{Z.-Y.} \bibnamefont{Ou}}, \bibnamefont{and}
  \bibinfo{author}{\bibfnamefont{L.}~\bibnamefont{Mandel}},
  \bibinfo{journal}{Physical Review Letters} \textbf{\bibinfo{volume}{59}},
  \bibinfo{pages}{2044} (\bibinfo{year}{1987}).

\bibitem[{\citenamefont{Gru{\ss}mayer and Herten}(2014)}]{grussmayer2014photon}
\bibinfo{author}{\bibfnamefont{K.~S.} \bibnamefont{Gru{\ss}mayer}}
  \bibnamefont{and} \bibinfo{author}{\bibfnamefont{D.-P.}
  \bibnamefont{Herten}}, in \emph{\bibinfo{booktitle}{Advanced Photon
  Counting}} (\bibinfo{publisher}{Springer}, \bibinfo{year}{2014}), pp.
  \bibinfo{pages}{159--190}.

\bibitem[{\citenamefont{Schwartz et~al.}(2013)\citenamefont{Schwartz, Levitt,
  Tenne, Itzhakov, Deutsch, and Oron}}]{schwartz2013superresolution}
\bibinfo{author}{\bibfnamefont{O.}~\bibnamefont{Schwartz}},
  \bibinfo{author}{\bibfnamefont{J.~M.} \bibnamefont{Levitt}},
  \bibinfo{author}{\bibfnamefont{R.}~\bibnamefont{Tenne}},
  \bibinfo{author}{\bibfnamefont{S.}~\bibnamefont{Itzhakov}},
  \bibinfo{author}{\bibfnamefont{Z.}~\bibnamefont{Deutsch}}, \bibnamefont{and}
  \bibinfo{author}{\bibfnamefont{D.}~\bibnamefont{Oron}},
  \bibinfo{journal}{Nano Letters} \textbf{\bibinfo{volume}{13}},
  \bibinfo{pages}{5832} (\bibinfo{year}{2013}).

\bibitem[{\citenamefont{Leung and Chou}(2011)}]{leung2011review}
\bibinfo{author}{\bibfnamefont{B.~O.} \bibnamefont{Leung}} \bibnamefont{and}
  \bibinfo{author}{\bibfnamefont{K.~C.} \bibnamefont{Chou}},
  \bibinfo{journal}{Applied Spectroscopy} \textbf{\bibinfo{volume}{65}},
  \bibinfo{pages}{967} (\bibinfo{year}{2011}).

\bibitem[{\citenamefont{Zurek}(2003)}]{zurek2003decoherence}
\bibinfo{author}{\bibfnamefont{W.~H.} \bibnamefont{Zurek}},
  \bibinfo{journal}{Reviews of Modern Physics} \textbf{\bibinfo{volume}{75}},
  \bibinfo{pages}{715} (\bibinfo{year}{2003}).

\bibitem[{\citenamefont{Zurek}(2009)}]{zurek2009quantum}
\bibinfo{author}{\bibfnamefont{W.~H.} \bibnamefont{Zurek}},
  \bibinfo{journal}{Nature Physics} \textbf{\bibinfo{volume}{5}},
  \bibinfo{pages}{181} (\bibinfo{year}{2009}).

\bibitem[{\citenamefont{Kumar}(1990)}]{kumar1990quantum}
\bibinfo{author}{\bibfnamefont{P.}~\bibnamefont{Kumar}},
  \bibinfo{journal}{Optics Letters} \textbf{\bibinfo{volume}{15}},
  \bibinfo{pages}{1476} (\bibinfo{year}{1990}).

\bibitem[{\citenamefont{Parameswaran et~al.}(2002)\citenamefont{Parameswaran,
  Route, Kurz, Roussev, Fejer, and Fujimura}}]{parameswaran2002highly}
\bibinfo{author}{\bibfnamefont{K.~R.} \bibnamefont{Parameswaran}},
  \bibinfo{author}{\bibfnamefont{R.~K.} \bibnamefont{Route}},
  \bibinfo{author}{\bibfnamefont{J.~R.} \bibnamefont{Kurz}},
  \bibinfo{author}{\bibfnamefont{R.~V.} \bibnamefont{Roussev}},
  \bibinfo{author}{\bibfnamefont{M.~M.} \bibnamefont{Fejer}}, \bibnamefont{and}
  \bibinfo{author}{\bibfnamefont{M.}~\bibnamefont{Fujimura}},
  \bibinfo{journal}{Optics Letters} \textbf{\bibinfo{volume}{27}},
  \bibinfo{pages}{179} (\bibinfo{year}{2002}).

\bibitem[{\citenamefont{Ikuta et~al.}(2011)\citenamefont{Ikuta, Kusaka, Kitano,
  Kato, Yamamoto, Koashi, and Imoto}}]{ikuta2011wide}
\bibinfo{author}{\bibfnamefont{R.}~\bibnamefont{Ikuta}},
  \bibinfo{author}{\bibfnamefont{Y.}~\bibnamefont{Kusaka}},
  \bibinfo{author}{\bibfnamefont{T.}~\bibnamefont{Kitano}},
  \bibinfo{author}{\bibfnamefont{H.}~\bibnamefont{Kato}},
  \bibinfo{author}{\bibfnamefont{T.}~\bibnamefont{Yamamoto}},
  \bibinfo{author}{\bibfnamefont{M.}~\bibnamefont{Koashi}}, \bibnamefont{and}
  \bibinfo{author}{\bibfnamefont{N.}~\bibnamefont{Imoto}},
  \bibinfo{journal}{Nature Communications} \textbf{\bibinfo{volume}{2}},
  \bibinfo{pages}{1544} (\bibinfo{year}{2011}).

\bibitem[{\citenamefont{De~Greve et~al.}(2012)\citenamefont{De~Greve, Yu,
  McMahon, Pelc, Natarajan, Kim, Abe, Maier, Schneider, Kamp
  et~al.}}]{de2012quantum}
\bibinfo{author}{\bibfnamefont{K.}~\bibnamefont{De~Greve}},
  \bibinfo{author}{\bibfnamefont{L.}~\bibnamefont{Yu}},
  \bibinfo{author}{\bibfnamefont{P.~L.} \bibnamefont{McMahon}},
  \bibinfo{author}{\bibfnamefont{J.~S.} \bibnamefont{Pelc}},
  \bibinfo{author}{\bibfnamefont{C.~M.} \bibnamefont{Natarajan}},
  \bibinfo{author}{\bibfnamefont{N.~Y.} \bibnamefont{Kim}},
  \bibinfo{author}{\bibfnamefont{E.}~\bibnamefont{Abe}},
  \bibinfo{author}{\bibfnamefont{S.}~\bibnamefont{Maier}},
  \bibinfo{author}{\bibfnamefont{C.}~\bibnamefont{Schneider}},
  \bibinfo{author}{\bibfnamefont{M.}~\bibnamefont{Kamp}}, \bibnamefont{et~al.},
  \bibinfo{journal}{Nature} \textbf{\bibinfo{volume}{491}},
  \bibinfo{pages}{421} (\bibinfo{year}{2012}).

\bibitem[{\citenamefont{Chou et~al.}(1998)\citenamefont{Chou, Hauden, Arbore,
  and Fejer}}]{chou19981}
\bibinfo{author}{\bibfnamefont{M.}~\bibnamefont{Chou}},
  \bibinfo{author}{\bibfnamefont{J.}~\bibnamefont{Hauden}},
  \bibinfo{author}{\bibfnamefont{M.}~\bibnamefont{Arbore}}, \bibnamefont{and}
  \bibinfo{author}{\bibfnamefont{M.}~\bibnamefont{Fejer}},
  \bibinfo{journal}{Optics Letters} \textbf{\bibinfo{volume}{23}},
  \bibinfo{pages}{1004} (\bibinfo{year}{1998}).

\bibitem[{\citenamefont{Vandevender and Kwiat}(2004)}]{vandevender2004high}
\bibinfo{author}{\bibfnamefont{A.~P.} \bibnamefont{Vandevender}}
  \bibnamefont{and} \bibinfo{author}{\bibfnamefont{P.~G.} \bibnamefont{Kwiat}},
  \bibinfo{journal}{Journal of Modern Optics} \textbf{\bibinfo{volume}{51}},
  \bibinfo{pages}{1433} (\bibinfo{year}{2004}).

\bibitem[{\citenamefont{Ma et~al.}(2009)\citenamefont{Ma, Slattery, and
  Tang}}]{ma2009experimental}
\bibinfo{author}{\bibfnamefont{L.}~\bibnamefont{Ma}},
  \bibinfo{author}{\bibfnamefont{O.}~\bibnamefont{Slattery}}, \bibnamefont{and}
  \bibinfo{author}{\bibfnamefont{X.}~\bibnamefont{Tang}},
  \bibinfo{journal}{Optics Express} \textbf{\bibinfo{volume}{17}},
  \bibinfo{pages}{14395} (\bibinfo{year}{2009}).

\end{thebibliography}
\begin{center}
\section*{\large Supplemental Materials}
\end{center}

\section*{Theoretical Overview}

First, we review the mathematics behind color erasure detectors~\cite{cotler2016entanglement}.  Suppose we have two types of photons $\gamma_1$ and $\gamma_2$, where the wavelength of $\gamma_2$ is shorter than that of $\gamma_1$.
We also consider a third wavelength $\gamma_3$ whose energy is the difference in energies between $\gamma_2$ and $\gamma_1$.  We prepare a coherent state of $\gamma_3$ photons, denoted by
\begin{equation}
|\alpha, \text{coh.}\rangle_{\gamma_3} := e^{-\frac{|\alpha|^2}{2}} \sum_{n=0}^\infty \frac{\alpha^n}{\sqrt{n!}} \, |n\rangle_{\gamma_3}\,,
\end{equation}
where $\alpha$ is a complex number, and $|n\rangle_{\gamma_3}$ is a number state with $n$ $\gamma_3$ photons in an incoming mode.  The average number of photons in the state $|\alpha, \text{coh.}\rangle_{\gamma_3}$ is $\!\,_{\gamma_3}\langle \alpha, \text{coh.}| \,\widehat{n} \,|\alpha, \text{coh.}\rangle_{\gamma_3} = |\alpha|^2$.  In our setting, the average number of photons $|\alpha|^2$ is large, which holds when the pump is a strong laser.

Our PPLN will have an input state either of the form
\begin{equation}
|1\rangle_{\gamma_1} \otimes |0\rangle_{\gamma_2} \otimes |\alpha, \text{coh.}\rangle_{\gamma_3}
\end{equation}
which has a single $\gamma_1$ photon and a coherent state of $\gamma_3$ photons, or
\begin{equation}
|0\rangle_{\gamma_1} \otimes |1\rangle_{\gamma_2} \otimes |\alpha, \text{coh.}\rangle_{\gamma_3}
\end{equation}
which has a single $\gamma_2$ photon and a coherent state of $\gamma_3$ photons.  The input state evolves with the Hamiltonian~\cite{kumar1990quantum}
\begin{equation}
H = i \chi \left(a_{\gamma_1} \otimes a_{\gamma_2}^\dagger \otimes a_{\gamma_3} - a_{\gamma_1}^\dagger \otimes a_{\gamma_2} \otimes a_{\gamma_3}^\dagger \right)
\end{equation}
where $a^\dagger, a$ are creation and annihilation operators.  Evolving for a time $T$, we have:
\begin{align}
& e^{- i H T}  |1\rangle_{\gamma_1} \otimes |0\rangle_{\gamma_2} \otimes |\alpha, \text{coh.}\rangle_{\gamma_3} \nonumber \\
& \quad =  |1\rangle_{\gamma_1} \otimes |0\rangle_{\gamma_2} \otimes \cos\big(\chi \,T \sqrt{\widehat{n}_{\gamma_3}}\big)  |\alpha, \text{coh.}\rangle_{\gamma_3} + |0\rangle_{\gamma_1} \otimes |1\rangle_{\gamma_2} \otimes a_{\gamma_3} \, \frac{\sin\big(\chi \,T \, \sqrt{\widehat{n}_{\gamma_3}}\big)}{\sqrt{\widehat{n}_{\gamma_3}}}  |\alpha, \text{coh.}\rangle_{\gamma_3} \\
&e^{- i H T}  |0\rangle_{\gamma_1} \otimes |1\rangle_{\gamma_2} \otimes |\alpha, \text{coh.}\rangle_{\gamma_3} \nonumber \\
& \quad =  - |1\rangle_{\gamma_1} \otimes |0\rangle_{\gamma_2} \otimes a_{\gamma_3}^\dagger \, \frac{\sin\big(\chi \,T \, \sqrt{\widehat{n}_{\gamma_3}+1}\big)}{\sqrt{\widehat{n}_{\gamma_3}+1}}   |\alpha, \text{coh.}\rangle_{\gamma_3} + |0\rangle_{\gamma_1} \otimes |1\rangle_{\gamma_2} \otimes \cos\big(\chi \,T \sqrt{\widehat{n}_{\gamma_3}+1}\big)  |\alpha, \text{coh.}\rangle_{\gamma_3}
\end{align}
Let us call the first state $|\Psi_1\rangle_{\gamma_1 \, \gamma_2 \, \gamma_3}$ and the second state $|\Psi_2\rangle_{\gamma_1 \, \gamma_2 \, \gamma_3}$.  Notice that both states are entangled between the $\gamma_1$, $\gamma_2$ modes, and the $\gamma_3$ mode.  The first state is a superposition of two possibilities: (i) the $\gamma_1$ remains, and (ii) the $\gamma_1$ absorbs a $\gamma_3$ and is upconverted to a $\gamma_2$.  The second state is similarly a superposition of the two possibilities: (i) the $\gamma_2$ remains, and (ii) the $\gamma_2$ emits a $\gamma_3$ photon and is downconverted to a $\gamma_1$.

Next, we put the state through a filter which only lets through $\gamma_2$ photons, and proceed if a $\gamma_2$ photon is outputted.  This is equivalent to post-selecting by projecting onto $\textbf{1} \otimes |1\rangle_{\gamma_2}\,_{\gamma_2}\langle 1| \otimes \textbf{1}$ and renormalizing the residual state, as
\begin{align}
|\widetilde{\Psi}_1\rangle_{\gamma_1 \, \gamma_2 \, \gamma_3} &= \frac{\big(\textbf{1} \otimes |1\rangle_{\gamma_2}\,_{\gamma_2}\langle 1| \otimes \textbf{1} \big) |\Psi_1\rangle_{\gamma_1 \,\gamma_2 \,\gamma_3}}{\sqrt{\,_{\gamma_1 \, \gamma_2 \, \gamma_3}\langle \Psi_1|  \big(\textbf{1} \otimes |1\rangle_{\gamma_2} \!\,_{\gamma_2}\langle 1 | \otimes \textbf{1} \big) |\Psi_1\rangle_{\gamma_1 \, \gamma_2 \, \gamma_3}}} \\ \nonumber \\
|\widetilde{\Psi}_2\rangle_{\gamma_1 \, \gamma_2 \, \gamma_3} &= \frac{\big(\textbf{1} \otimes |1\rangle_{\gamma_2}\,_{\gamma_2}\langle 1| \otimes \textbf{1} \big) |\Psi_2\rangle_{\gamma_1 \,\gamma_2 \,\gamma_3}}{\sqrt{\,_{\gamma_1 \, \gamma_2 \, \gamma_3}\langle \Psi_2|  \big(\textbf{1} \otimes |1\rangle_{\gamma_2} \!\,_{\gamma_2}\langle 1 | \otimes \textbf{1} \big) |\Psi_2\rangle_{\gamma_1 \, \gamma_2 \, \gamma_3}}}\,.
\end{align}
Letting $\alpha = e^{i \phi} \, \sqrt{N}$ for $N$ large, we find that
\begin{equation}
\left|\!\,_{\gamma_1 \, \gamma_2 \, \gamma_3}\langle \widetilde{\Psi}_1|\widetilde{\Psi}_2\rangle_{\gamma_1 \, \gamma_2 \, \gamma_3} \right| = 1 \,\,+ \mathcal{O}(1/\sqrt{N})
\end{equation}
and so $|\widetilde{\Psi}_1\rangle_{\gamma_1 \, \gamma_2 \, \gamma_3}$ and $|\widetilde{\Psi}_2\rangle_{\gamma_1 \, \gamma_2 \, \gamma_3}$ cannot be distinguished.  In plainer terms, this means if we evolve either $\gamma_1$ \textit{or} $\gamma_2$ in the PPLN with the $\gamma_3$ coherent state and then measure a $\gamma_2$ as the output, then the apparatus fundamentally cannot tell us whether $\gamma_2$ was originally a $\gamma_1$ or a $\gamma_2$.  Hence, the detection apparatus can erase the color information.

There is another, more illuminating way of reprocessing the above analysis.  Consider again the evolved states $|\Psi_1\rangle_{\gamma_1 \, \gamma_2 \, \gamma_3}$ and $|\Psi_2\rangle_{\gamma_1 \, \gamma_2 \, \gamma_3}$.  We can write down the corresponding density matrices
\begin{align}
\rho_{\gamma_1 \, \gamma_2 \, \gamma_3} = \big(|\Psi_1\rangle_{\gamma_1 \, \gamma_2 \, \gamma_3}\big) \big( \,_{\gamma_1 \, \gamma_2 \, \gamma_3}\langle \Psi_1| \big) \\
\sigma_{\gamma_1 \, \gamma_2 \, \gamma_3} = \big(|\Psi_2\rangle_{\gamma_1 \, \gamma_2 \, \gamma_2}\big) \big( \,_{\gamma_1 \, \gamma_2 \, \gamma_3}\langle \Psi_2| \big)
\end{align}
and trace out the $\gamma_3$ photons to obtain
\begin{align}
\rho_{\gamma_1 \, \gamma_2} &= \text{tr}_{\gamma_3}\left( \rho_{\gamma_1 \, \gamma_2 \, \gamma_3} \right) \\
\sigma_{\gamma_1 \, \gamma_2} &= \text{tr}_{\gamma_3}\left( \sigma_{\gamma_1 \, \gamma_2 \, \gamma_3} \right) \,.
\end{align}
Again letting $\alpha = e^{i \phi} \sqrt{N}$ and taking $N$ large, we can use the Euler-Maclaurin formula and a saddle point approximation to compute the explicit expressions of $\rho_{\gamma_1 \, \gamma_2}$ and $\sigma_{\gamma_1 \, \gamma_2}$.  We find that
\begin{align}
\rho_{\gamma_1 \, \gamma_2} &= \big( |\Phi_1 \rangle_{\gamma_1 \, \gamma_2} \big)\big(\,_{\gamma_1 \, \gamma_2}\langle \Phi_1 |\big) \\
\sigma_{\gamma_1 \, \gamma_2} &= \big( |\Phi_2 \rangle_{\gamma_1 \, \gamma_2} \big)\big(\,_{\gamma_1 \, \gamma_2}\langle \Phi_2 |\big)
\end{align}
where
\begin{align}
\label{Phi1eq1}
|\Phi_1\rangle_{\gamma_1 \, \gamma_2} &=  \cos( \chi T \sqrt{N}) \, |1\rangle_{\gamma_1} \otimes |0\rangle_{\gamma_2}  + e^{i \phi}\,\sin(\chi T \sqrt{N}) \, |0\rangle_{\gamma_1} \otimes |1\rangle_{\gamma_2}  \\
\label{Phi2eq1}
|\Phi_2\rangle_{\gamma_1 \, \gamma_2} &=  - e^{- i \phi} \sin( \chi T \sqrt{N}) \, |1\rangle_{\gamma_1} \otimes |0\rangle_{\gamma_2}  + \cos(\chi T \sqrt{N}) \, |0\rangle_{\gamma_1} \otimes |1\rangle_{\gamma_2}
\end{align}
up to $\mathcal{O}(1/\sqrt{N})$ corrections.  Notice that $|\Phi_1\rangle_{\gamma_1 \, \gamma_2}$ and $|\Phi_2\rangle_{\gamma_1 \, \gamma_2}$ are mutually orthogonal.  From the above equations, we see that the PPLN performs a rotation in color space.  It is clear that projecting either $|\Phi_1\rangle_{\gamma_1 \, \gamma_2} $ or $|\Phi_2\rangle_{\gamma_1 \, \gamma_2} $ onto the $|0\rangle_{\gamma_1} \otimes |1\rangle_{\gamma_2}$ state and renormalizing will yield $|0\rangle_{\gamma_1} \otimes |1\rangle_{\gamma_2}$, and so we see more simply that the detector is blind to the initial color of the photon.

The key observation is that in the large photon limit of the coherent state, we can truly treat $\gamma_3$ as a classical light field which is incapable of recording information about individual photons.  In fact, we see from Eqn.'s~\eqref{Phi1eq1} and~\eqref{Phi2eq1} that the \textit{effective} Hamiltonian which evolves the $\gamma_1$, $\gamma_2$ modes is
\begin{equation}
H_{\text{eff}} = i \chi \sqrt{N} \left(e^{i \phi}\, a_{\gamma_1} \otimes a_{\gamma_2}^\dagger - e^{- i \phi}\,a_{\gamma_1}^\dagger \otimes a_{\gamma_2} \right)
\end{equation}
which clearly performs a rotation in color space.

To perform chromatic intensity interferometry, we consider two sources $1$ and $2$, emitting $\gamma_1$ and $\gamma_2$ photons, respectively.  We also have two color erasure detectors $A$ and $B$.  Let $D_{1A}$ be the probability amplitude that a single photon emitted from $1$ is received by $A$.  The probability amplitudes $D_{2A}$, $D_{1B}$, $D_{2B}$ are defined similarly.  If a single photon is received by each detector (this is a form of post-selection), then we have the state
\begin{equation}
\label{receivedState1}
D_{1A} D_{2B} |1\rangle_{\gamma_1, A} \otimes |0\rangle_{\gamma_2, A} \otimes |0\rangle_{\gamma_1, B} \otimes |1\rangle_{\gamma_2, B} + D_{1B} D_{2A} |0\rangle_{\gamma_1, A} \otimes |1\rangle_{\gamma_2, A} \otimes |1\rangle_{\gamma_1, B} \otimes |0\rangle_{\gamma_2, B}
\end{equation}
where the first term corresponds to having a $\gamma_1$ photon at $A$ and a $\gamma_2$ photon at $B$, and the second term corresponds to having a $\gamma_2$ photon at $B$ and a $\gamma_1$ photon at $A$.  After the detectors $A$ and $B$ process their photons and we post-select on $A$ outputting $\gamma_2$ and $B$ outputting $\gamma_2$, we are left with the state
\begin{equation}
e^{i \phi} \cos(\chi T \sqrt{N}) \sin(\chi T \sqrt{N}) (D_{1A} D_{2B} + D_{1B} D_{2A}) |0\rangle_{\gamma_1, A} \otimes |1\rangle_{\gamma_2, A} \otimes |0\rangle_{\gamma_1, B} \otimes |1\rangle_{\gamma_2, B}
\end{equation}
with probability
\begin{align}
\label{HBTinterfere1}
&\cos^2(\chi T \sqrt{N}) \sin^2(\chi T \sqrt{N}) \,|D_{1A} D_{2B} + D_{1B} D_{2A}|^2 \nonumber \\
=\, & \cos^2(\chi T \sqrt{N}) \sin^2(\chi T \sqrt{N})  \, \bigg(|D_{1A} D_{2B}|^2 + |D_{1B} D_{2A}|^2 + 2 \, \text{Re}\left(D_{1A} D_{2B} D_{1B}^* D_{2A}^* \right) \bigg)
\end{align}
which contains the Hanbury-Brown Twiss interference term~\cite{mandel1995optical, baym1998physics, cotler2016entanglement} $\text{Re}\left(D_{1A} D_{2B} D_{1B}^* D_{2A}^* \right)$.

As a concrete example, suppose that $\chi T \sqrt{N} = \pi/4$, and that the length from $1$ to $A$ is $L_{1A}$.  The lengths $L_{2A}$, $L_{1B}$, $L_{2B}$ are defined similarly.  Denoting the wavelengths of $\gamma_1$ and $\gamma_2$ by $\lambda_1$ and $\lambda_2$ respectively, and assuming the sources $1$ and $2$ emit photons with equal probability, we have
\begin{align}
\label{Damplitudes1}
D_{1A} &= \frac{1}{\sqrt{2}} \, e^{i\, 2\pi \, L_{1A}/\lambda_1 + i \, \theta_1}\,, \quad D_{1B} = \frac{1}{\sqrt{2}} \, e^{i\,2\pi\, L_{1B}/\lambda_1 + i \, \theta_1}\,, \nonumber \\
D_{2A} &= \frac{1}{\sqrt{2}} \, e^{i\, 2\pi\,L_{2A}/\lambda_2 + i \, \theta_2}\,, \quad  D_{2B} = \frac{1}{\sqrt{2}} \, e^{i\, 2\pi\,L_{2B}/\lambda_2 + i \, \theta_2}\,,
\end{align}
where $\theta_1, \theta_2$ are phases associated with the emission of photons from sources $1$ and $2$, respectively.  In this case, Eqn.~\eqref{HBTinterfere1} becomes
\begin{equation}
\frac{1}{8} \left[1 + \cos\left(2 \pi \left(\frac{L_{1A}}{\lambda_1} + \frac{L_{2B}}{\lambda_2} - \frac{L_{1B}}{\lambda_1} - \frac{L_{2A}}{\lambda_2}\right) \right) \right]
\end{equation}
where the interference term $\text{Re}\left(D_{1A} D_{2B} D_{1B}^* D_{2A}^* \right)$ is $\frac{1}{4}\,\cos\left(2 \pi \left(\frac{L_{1A}}{\lambda_1} + \frac{L_{2B}}{\lambda_2} - \frac{L_{1B}}{\lambda_1} - \frac{L_{2A}}{\lambda_2} \right) \right)$.  Note that the interference term is independent of $\theta_1$ and $\theta_2$.  Accordingly, we can achieve interference between two mutually incoherent sources~\cite{baym1998physics, cotler2016entanglement}.  In this case, $\theta_1$ and $\theta_2$ may be strongly time-dependent, but nonetheless cancel out in the interference term.

In the analysis above, we have assumed that our photon sources each emit exactly one photon within some time window.  We can relax this assumption in various ways.  First, suppose that the first and second sources emit coherent superpositions of photon number states, namely
\begin{align}
\label{coherentemit1}
& c_0 \, |0\rangle_{\gamma_1} + c_1 \, |1\rangle_{\gamma_1} + c_2 \, |2\rangle_{\gamma_1} + \cdots \\
\label{coherentemit2}
& d_0 \, |0\rangle_{\gamma_2} + d_1 \, |1\rangle_{\gamma_2} + d_2 \, |2\rangle_{\gamma_2} + \cdots\,,
\end{align}
respectively, where the $c_i$'s and $d_i$'s are complex numbers satisfying $\sum_i |c_i|^2 = \sum_i |d_i|^2 = 1$.  Then detectors $A$ and $B$ receive the state
\begin{align}
&c_1 \, d_1 \, \bigg(D_{1A} D_{2B} |1\rangle_{\gamma_1, A} \otimes |0\rangle_{\gamma_2, A} \otimes |0\rangle_{\gamma_1, B} \otimes |1\rangle_{\gamma_2, B} + D_{1B} D_{2A} |0\rangle_{\gamma_1, A} \otimes |1\rangle_{\gamma_2, A} \otimes |1\rangle_{\gamma_1, B} \otimes |0\rangle_{\gamma_2, B} \bigg)\nonumber \\
+ \, & c_2 \, d_0 \, D_{1A} D_{1B} |1\rangle_{\gamma_1, A} \otimes |0\rangle_{\gamma_2, A} \otimes |1\rangle_{\gamma_1, B} \otimes |0\rangle_{\gamma_2, B} + c_2 \, d_0 \, D_{2A} D_{2B} |0\rangle_{\gamma_1, A} \otimes |1\rangle_{\gamma_2, A} \otimes |0\rangle_{\gamma_1, B} \otimes |1\rangle_{\gamma_2, B} \nonumber \\
+ \, & \cdots
\end{align}
The first line of the above equation corresponds to each source emitting a single photon and each detector receiving a single photon, and so has the same form as Eqn.~\eqref{receivedState1}.  The second line corresponds to (i) the first source emitting two photons and the second source emitting no photons, and each detector receiving a single photon, and (ii) the first source emitting no photons and the second source emitting two photons, and each detector receiving a single photon.  The final line with the ellipses accounts for the remaining terms.

As before, after the detectors $A$ and $B$ process their photons and we post-select on $A$ outputting a single $\gamma_2$ and $B$ outputting a single $\gamma_2$, we obtain the state
\begin{align}
&\bigg( c_1 \, d_1 \, e^{i \phi} \cos(\chi T \sqrt{N}) \, \sin(\chi T \sqrt{N} )\,(D_{1A} D_{2B}  + D_{1B} D_{2A}) + c_2 \, d_0 \, e^{2 i \phi} \sin^2(\chi T \sqrt{N}) \, D_{1A} D_{1B} \nonumber \\
& \qquad \qquad \qquad \qquad \qquad \quad + c_0 \, d_2 \, \cos^2(\chi T \sqrt{N}) D_{2A} D_{2B} \bigg) |0\rangle_{\gamma_1, A} \otimes |1\rangle_{\gamma_2, A} \otimes |0\rangle_{\gamma_1, B} \otimes |1\rangle_{\gamma_2, B}
\end{align}
with probability
\begin{align}
\label{HBTinterfere2}
& |c_1|^2 |d_1|^2 \cos^2(\chi T \sqrt{N}) \sin^2(\chi T \sqrt{N}) \, |D_{1A} D_{2B} + D_{1B} D_{2A}|^2  \nonumber \\
+ \,\,& |c_2|^2 |d_0|^2 \sin^4(\chi T \sqrt{N}) \, |D_{1A} D_{1B}|^2 + |c_0|^2 |d_2|^2 \cos^4(\chi T \sqrt{N}) \, |D_{2A} D_{2B}|^2  \nonumber \\
+ \,\,&  2 \cos(\chi T \sqrt{N}) \sin^3(\chi T \sqrt{N})\, \text{Re}\left(c_1 \, d_1\, c_2^*\,d_0^* \,e^{- i \phi} (D_{1A} D_{2B} + D_{1B} D_{2A}) \, D_{1A}^* D_{1B}^* \right) \nonumber \\
+ \,\,&  2 \cos^3(\chi T \sqrt{N}) \sin(\chi T \sqrt{N})\, \text{Re}\left(c_1 \, d_1\, c_0^*\,d_2^* \,e^{i \phi} (D_{1A} D_{2B} + D_{1B} D_{2A}) \, D_{2A}^* D_{2B}^* \right) \nonumber \\
+ \,\,&  2 \cos^2(\chi T \sqrt{N}) \sin^2(\chi T \sqrt{N})\, \text{Re}\left(c_2 \, d_0\, c_0^*\,d_2^* \,e^{2 i \phi} \, D_{1A} D_{1B} D_{2A}^* D_{2B}^* \right)\,.
\end{align}
Several remarks are in order.  First, notice that if $|c_0| |d_2| \ll |c_1| |d_1|$ and $|c_2| |d_0| \ll |c_1| |d_1|$, then the first term in the above equation dominates, which recovers the same interference as in Eqn.~\eqref{HBTinterfere1}.

Now suppose $D_{1A}, D_{2A}, D_{1B}, D_{2B}$ are the same as in Eqn.~\eqref{Damplitudes1}, but with $\theta_1 = \theta_1(t)$ and $\theta_2 = \theta_2(t)$ time-dependent and rapidly changing faster than the timescale of photon emission from the sources.  It is natural to assume that $\theta_1(t)$ and $\theta_2(t)$ are each ergodic on $[0, 2\pi]$.  In this case, it is easy to check that the time average of Eqn.~\eqref{HBTinterfere2}  is simply the first two lines of the equation, i.e.\! the last three lines vanish under time averaging.  This is because only the terms in the first two lines are independent of $\theta_1(t)$ and $\theta_2(t)$, whereas the remaining terms do depend on $\theta_1(t)$ and $\theta_2(t)$ and so average to zero.  Accordingly, we can still recover the desired Hanbury-Brown Twiss interference term contained in $|D_{1A} D_{2B} + D_{1B} D_{2A}|^2$ since this is the only remaining term sensitive to relative phases between the photon probability amplitudes after time-averaging.  Note that the experimental collection of data automatically incorporates time-averaging, since one averages results over many trials.

Instead of requiring the sources to emit coherent superpositions of photon number states as in Eqn.~\eqref{coherentemit1} and~\eqref{coherentemit2}, we can also accommodate for arbitrarily incoherent density matrices of photon states.  For instance, suppose that the first and second sources emit completely incoherent sums of photon number states described by the density matrices
\begin{align}
p_0 \, |0\rangle \langle 0|_{\gamma_1} + p_1 \, |1\rangle \langle 1|_{\gamma_1}  + p_2 \, |2\rangle \langle 2|_{\gamma_1}  + \cdots \\
q_0 \, |0\rangle \langle 0|_{\gamma_2} + q_1 \, |1\rangle \langle 1|_{\gamma_2}  + q_2 \, |2\rangle \langle 2|_{\gamma_2}  + \cdots
\end{align}
This occurs, for instance, if the sources are thermally populating the photon modes, and $\{p_i\}, \{q_i\}$ are classical Gibbs distributions.  Running through the same analysis as above, after the detectors process their photons and we post-select on $A$ outputting a single $\gamma_2$ and $B$ outputting a single $\gamma_2$, we are left with the density matrix
\begin{align}
& \bigg(p_1 \, q_1 \, \cos^2(\chi T \sqrt{N}) \sin^2(\chi T \sqrt{N}) \, |D_{1A} D_{2B} + D_{1B} D_{2A}|^2 + p_2 \, q_0 \sin^4(\chi T \sqrt{N}) \, |D_{1A} D_{1B}|^2 \nonumber \\
& \qquad \qquad +  p_0 \, q_2\, \cos^4(\chi T \sqrt{N}) \, |D_{2A} D_{2B}|^2 \bigg) \, |0\rangle \langle 0|_{\gamma_1, A} \otimes |1\rangle \langle 1|_{\gamma_2, A} \otimes |0\rangle \langle 0|_{\gamma_1, B} \otimes |1\rangle \langle 1|_{\gamma_2, B}
\end{align}
with probability
\begin{align}
\label{HBTinterfere3}
& p_1 \, q_1 \, \cos^2(\chi T \sqrt{N}) \sin^2(\chi T \sqrt{N}) \, |D_{1A} D_{2B} + D_{1B} D_{2A}|^2  \nonumber \\
+ \,\,& p_2 \, q_0 \sin^4(\chi T \sqrt{N}) \, |D_{1A} D_{1B}|^2 + p_0 \, q_2\, \cos^4(\chi T \sqrt{N}) \, |D_{2A} D_{2B}|^2 \,.
\end{align}
Notice that this has a similar form as Eqn.~\eqref{HBTinterfere2}, but without the unwanted interference terms in the last three lines of Eqn.~\eqref{HBTinterfere2}.  As before, we recover the desired Hanbury-Brown Twiss interference term contained in $|D_{1A} D_{2B} + D_{1B} D_{2A}|^2$, which is in fact the only interference term in Eqn.~\eqref{HBTinterfere3}.

\section*{Experimental Methods}

\begin{figure*}[t]
\centering
\resizebox{14cm}{!}{\includegraphics{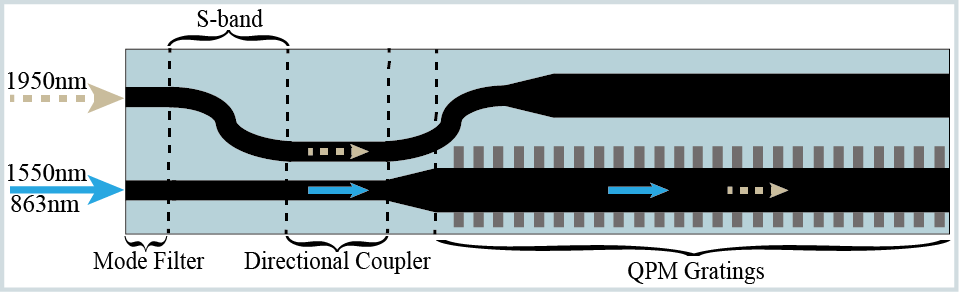}}
\caption{Setup for chromatic intensity interferometer for thermal light sources.}
\label{PPLN}
\end{figure*}

\textit{Experiment details for PPLN waveguide}
A key device within each color erasure detector is an integrated PPLN waveguide. We fabricated reverse-proton-exchange (RPE) PPLN waveguides ~\cite{parameswaran2002highly} with a total length of $52$ mm for both difference-frequency generation (DFG) ~\cite{ikuta2011wide,de2012quantum,chou19981} between $863$ nm light and the $1950$ nm pump, and sum-frequency generation (SFG) ~\cite{vandevender2004high,ma2009experimental,kumar1990quantum} between $1550$ nm light and the $1950$ nm pump. We use an integrated waveguide structure consisting of a bent waveguide and a straight waveguide with an entrance center-to-center separation of $127$ $\mu$m, as shown in Fig.\ref{PPLN}. The main features of the integrated structure are two $5.5$ $\mu$m wide mode filters, a directional coupler used as a wavelength combiner, and a $8$ $\mu$m wide uniform straight waveguide with $44$ mm long quasi-phase-matching (QPM) gratings for optical frequency nonlinear mixing. $1550$ nm photons and $863$ nm photons are combined by a $1550$ nm/$863$ nm wavelength-division multiplexer (WDM) before they enter the straight waveguide together. $1950$ nm photons enter the bent waveguide and pass through a $3.5$ mm long S-band before entering the directional coupler. With a waveguide width of $5.5$ $\mu$m, an edge-to-edge spacing of $5.5$ $\mu$m, and a length of $2.5$ mm, the directional coupler combines the $1950$ nm pump, the $863$ nm photons and the $1550$ nm photons into the same straight waveguide with negligible losses for both signals. The combined photons then enter the QPM mixing region which is poled with a period of $20$ $\mu$m. The input and output of the waveguides are fiber-pigtailed by two polarization maintaining (PM) taper-fibers and a PM $1550$ nm fiber, respectively. The total waveguide throughputs are $-3.5$ dB and $-4$ dB for $1550$ nm and $863$ nm, respectively.

\textit{Experiment details for coherent sources.} We use a 1550 nm laser as the $\gamma_{1}$ source, and a 1950 nm laser as the $\gamma_3$ source.  In our proof-of principle experiment, the $\gamma_{2}$ source is generated by up-conversion of a separate $\gamma_{1}$ source by the $\gamma_{3}$ source.

\begin{table*}[h!]
\centering
\caption{Key parameters for laser case}
\begin{tabular}{c|c}
\hline
Parameter & Value\\
\hline
Detection efficiency of UCSPD & $19.5\%$\\
Wavelength of $\gamma_1$ source & 1549.800 nm\\
Wavelength of $\gamma_2$ source & 863.344 nm\\
Wavelength of $\gamma_3$ source & 1949.157 nm\\
Temperature of PPLN waveguide A& $36.4\,^{\circ}C$\\
Temperature of PPLN waveguide B& $52.9\,^{\circ}C$\\
Power of pump& $152.6$ mw\\
\hline
\end{tabular}
\label{tab:laserparameters}
\end{table*}

Our color erasure detector requires that $\nu_{\gamma_{1}}+\nu_{\gamma_{3}}=\nu_{\gamma_{2}}$, and that the photons involved be quasi-phase-matched in the PPLN waveguide.  We can adjust the temperature of the PPLN waveguide to change the refractive index so that quasi-phase-matching can be achieved, which can be diagnosed from the count rate of the detector. Our parameters, such as the photon wavelengths, temperature of PPLN waveguide, and pump power, are shown in Table I. Also note that the 1950 nm pumps for the two PPLN waveguides should be phase locked, to enable intensity interferometry.  In our experiment, the two 1950 nm pumps are siphoned from a single source and the phase noise is weaken by insulating fiber optical cables with cotton.

For chromatic interferometry, photons which do not participate in (partial) frequency conversion should be taken as noise or dark counts.  Only photons with polarization parallel to the optical axis can be up-converted or down-converted in the PPLN waveguide.  Fortunately, the PPLN waveguide does not let through photons with polarization vertical to the optical axis and thus imperfect polarization will not contribute to the dark counts.  We leverage this feature to control the number of photons received by each detector via polarization controllers after each light source.  Similarly, photons in higher-order spatial modes of the waveguide do not take part in the frequency-conversion process. Therefore, the coupled fiber should be matched to the lowest (radially symmetric) transverse mode at the input and output of PPLN waveguide.



\begin{figure*}[t]
\centering
\resizebox{14cm}{!}{\includegraphics{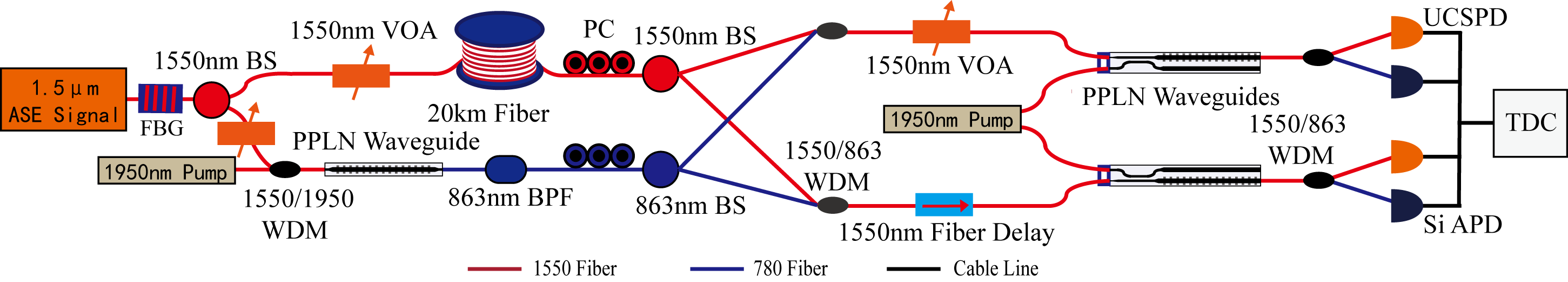}}
\caption{Setup for chromatic intensity interferometer for thermal light sources.}
\label{thermalsetup}
\end{figure*}

\begin{figure*}[t]
\centering
\resizebox{10cm}{!}{\includegraphics{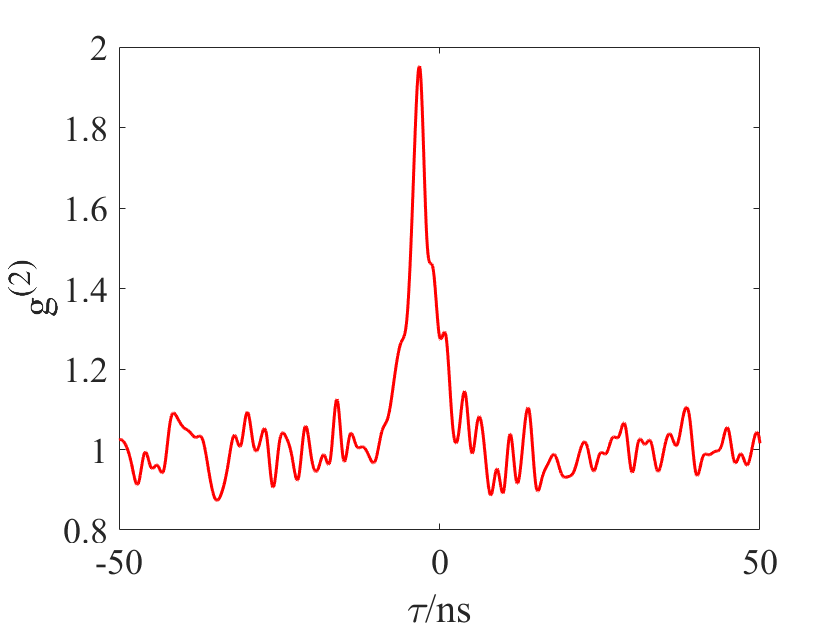}}
\vskip-.5cm
\caption{Second order coherence $g^{(2)}(\tau)$ of the thermal light source.}
\label{thermalg2}
\end{figure*}

\textit{Experiment details for thermal sources.}  Our setup for chromatic interferometry of thermal light sources is shown in Fig.\ref{thermalsetup}. To generate a thermal light source in the tabletop experiment, we use an ASE source with 30 nm spectral width and implement a 50 MHz bandwidth filter to select out one thermally populated mode. To test the photon number distribution of our thermal source, we pass the photons through a 50-50 beamsplitter, and record the arrival times at two detectors placed after each out port of the beamsplitter in order to calculate $g^{(2)}(\tau)$. As shown in Fig.\ref{thermalg2}, the $g^{(2)}(\tau)$ of the thermal source is approximately 2 within a coherence time.  In this setup, the $\gamma_1$ and $\gamma_2$ 

\begin{table*}[th!]
\vskip.5cm
\centering
\caption{Key parameters for thermal light source case}
\vskip.5cm
\begin{tabular}{c|c}
\hline
Parameter & Value\\
\hline
Detection efficiency of Si APD & $55\%$\\
Detection efficiency of UCSPD & $19.5\%$\\
Wavelength of $\gamma_1$ source & 1549.968 nm\\
Wavelength of $\gamma_2$ source & 863.396 nm\\
Wavelength of $\gamma_3$ source & 1949.157 nm\\
Temperature of PPLN waveguide A& $37.4\,^{\circ}C$\\
Temperature of PPLN waveguide B& $34.9\,^{\circ}C$\\
Power of pump& $192.3$ mw\\
Bandwidth of optical filter& $50$ MHz\\
\hline
\end{tabular}
\label{tab:thermalparameters}
\end{table*}

\noindent photons are generated from the same source. In order to destroy residual phase correlations between our $\gamma_1$ and $\gamma_2$ sources, we run the $\gamma_1$ photons through a 20 km spool of fiber.

We test chromatic interferometry in four scenarios: (i) both color erasure detectors output $\gamma_1$, (ii) detector $A$ outputs $\gamma_1$ and detector $B$ outputs $\gamma_2$, (iii) detector $A$ outputs $\gamma_2$ and detector $B$ outputs $\gamma_1$, and (iv) both detectors output $\gamma_2$.  The key parameters are shown in Table 2.

\textit{Calculating the second order coherence function.}
Let $I_A(t_1)$ be the intensity measured at detector $A$ at time $t_1$, and similarly let $I_B(t_2)$ be the intensity measured at detector $B$ at time $t_2$.  The definition of the second order coherence function $g^{(2)}(\tau)$ is:
\begin{equation}
g^{(2)}(\tau) = \frac{\int dt \, I_A(t) I_B(t + \tau)}{\left( \int dt \, I_A(t) \right) \left( \int dt \, I_B(t) \right)}
\end{equation}
where the suppressed limits of the integrals are limited by the length of our trials.  Note that in the limit of long integration time (i.e., the integrals $\int dt$ above are essentially $\int_{-\infty}^\infty dt$\,), we have
\begin{equation}
g^{(2)}(\tau = 0) = |D_{1A} D_{2B} + D_{1B} D_{2A}|^2
\end{equation}
in our notation from earlier.

In our experiment, we record the arrival time of each photon detected by a UCSPD.  To analysis the data, we set a 1 ns gate time and judge the coincidence of each count. When two counts from separate detectors fall into the same time bin, we register a coincidence.  Let $n_{\text{coincidence}}$ be the total number of coincidence counts over the course of a run, let $n_{\text{bin}}$ be the total number of time bins, and let $n_{A}$ and $n_{B}$ be the total counts of detector $A$ and detector $B$, respectively.   Then our calculation of $g^{(2)}(\tau)$ amounts to
\begin{equation}
g^{(2)}(\tau)=\frac{n_{\text{coincidence}} \cdot n_{\text{bin}}}{n_{A} \cdot n_{B}}
\end{equation}

\textit{Considerations for visibility of interference.}
In the theoretical overview, we wanted to choose $\theta = \chi T \sqrt{N} = \pi/4$ such that $\cos^2(\theta) = \sin^2(\theta) = 1/2$ to achieve maximum visibility of the interference.  In that analysis, we assumed that the same number of incoming $\gamma_1$ and $\gamma_2$ photons couple to the color erasure detectors.  However, in practice, the number of $\gamma_{1}$ and $\gamma_{2}$ photons which couple to a PPLN waveguide can be different.  Suppose we are post-selecting on $\gamma_1$.  Then letting $N_{\gamma_1}$ and $N_{\gamma_2}$ be the number of photons which couple to a PPLN waveguide over the course of an experimental run, to achieve maximal visibility we need to ensure that $N_{\gamma_{1}} \cos^{2}(\theta)=N_{\gamma_{2}} \sin^{2}(\theta)$ for both color erasure detectors $A$ and $B$.
\begin{figure*}[h!]
\centering
\resizebox{14cm}{!}{\includegraphics{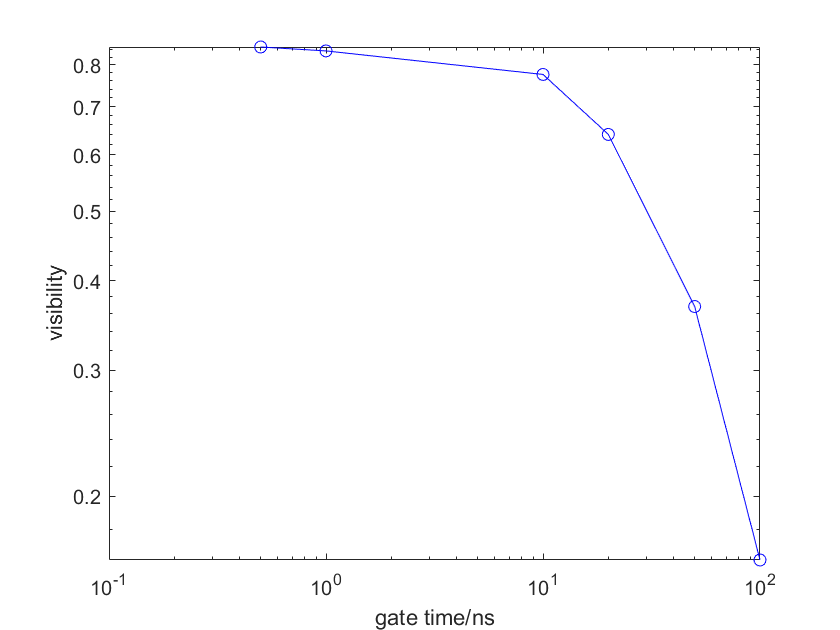}}
\caption{Relationship between visibility and gate time. The visibilty will not significantly increase when the gate time is less than 1 ns. Interference will begin to vanish when the gate time is more than 200 ns.}
\label{gatetime}
\end{figure*}
Another parameter that influences the visibility is the gate time of the coincidence counter, which corresponds to the time resolution of the detector. As shown in Fig.\ref{gatetime}, the visibility decreases as the gate time increases. For single photon detection, given the total count of each detector, increasing the gate time increases the probability that two coincident photons come from one light source, which will decrease the visibility of interference.  On the other hand, if the gate time is too small, then there will be fewer coincidence counts and hence a decreased signal-to-noise ratio.




  







\newpage
\end{document}